\documentclass[10pt, sigconf, nonacm]{acmart}
\AtBeginDocument{%
  }

\usepackage{algorithm}
\usepackage{graphicx}
\usepackage{textcomp}
\usepackage[dvipsnames]{xcolor}
\usepackage{booktabs}
\usepackage{soul}
\usepackage{tabularray}
\usepackage{multirow}
\usepackage{tabularray}
\usepackage{pifont}
\usepackage{wasysym}
\usepackage{hyperref}
\hypersetup{hidelinks}

\soulregister\cite7
\soulregister\autoref7

\begin{document}


\title{Clover: A Neural-Symbolic Agentic Harness with Stochastic Tree-of-Thoughts for Verified RTL Repair}

\newcommand{\rdba}{\emph{Clover}}
\newcommand{\circled}[1]{\raisebox{.5pt}{\textcircled{\raisebox{-.9pt} {#1}}}}


\author{Zizhang Luo}
\email{semiwaker@pku.edu.cn}
\orcid{0000-0002-7276-2317}
\affiliation{%
  \institution{Peking University}
  \city{Beijing}
  \country{China}
}

\author{Yansong Xu}
\email{yansongxu.mail@gmail.com}
\affiliation{%
  \institution{Peking University}
  \city{Beijing}
  \country{China}
}

\author{Runlin Guo}
\email{linyu22373@gmail.com}
\affiliation{%
  \institution{Peking University}
  \city{Beijing}
  \country{China}
}

\author{Fan Cui}
\email{pku_cf@stu.pku.edu.cn}
\orcid{0009-0007-0132-628X}
\affiliation{%
  \institution{Peking University}
  \city{Beijing}
  \country{China}
}

\author{Kexing Zhou}
\email{zhoukexing@pku.edu.cn}
\orcid{0000-0001-7635-3425}
\affiliation{%
  \institution{Peking University}
  \city{Beijing}
  \country{China}
}

\author{Mile Xia}
\email{milexia@stu.pku.edu.cn}
\affiliation{%
  \institution{Peking University}
  \city{Beijing}
  \country{China}
}

\author{Hongyuan Hou}
\email{houhy@stu.pku.edu.cn}
\affiliation{%
  \institution{Peking University}
  \city{Beijing}
  \country{China}
}

\author{Yuhao Luo}
\email{luoyuhao584@gmail.com}
\affiliation{%
  \institution{Peking University}
  \city{Beijing}
  \country{China}
}

\author{Yun Liang}
\email{ericlyun@pku.edu.cn}
\orcid{0000-0002-9076-7998}
\affiliation{%
  \institution{Peking University}
  \city{Beijing}
  \country{China}
}

\newcommand{\algorithmautorefname}{Algorithm}

\begin{abstract}

RTL program repair remains a critical bottleneck in hardware design and verification. Traditional automatic program repair (APR) methods rely on predefined templates and synthesis, limiting their bug coverage. Large language models (LLMs) and coding agents based on them offer flexibility but suffer from randomness and context corruption when handling long RTL code and waveforms.

We present Clover, a neural-symbolic agentic harness that orchestrates RTL repair as a structured search over code manipulations to explore a validated solution for the bug. Recognizing that different repair operations favor distinct strategies, Clover dynamically dispatches tasks to specialized LLM agents or symbolic solvers. At its core, Clover introduces stochastic tree-of-thoughts, a test-time scaling mechanism that manages the main agent’s context as a search tree, balancing exploration and exploitation for reliable outcomes. An RTL-specific toolbox further empowers agents to interact with the debugging environment.

Evaluated on the RTL‑repair benchmark, Clover fixes 96.8\% of bugs within a fixed time limit, covering 94\% and 63\% more bugs than both pure traditional and LLM‑based baselines, respectively, while achieving an average pass@1 rate of 87.5\%, demonstrating high reliability and effectiveness.

\end{abstract}

\maketitle

\section{Introduction}

\begin{figure}[t]
    \centering
    \includegraphics[width=\columnwidth]{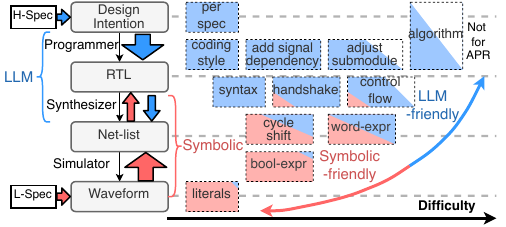}
    \caption{Spectrum of Program Repair Operations}
    \label{fig:category}
\end{figure}

Register‑transfer‑level (RTL) debugging is an indispensable task for hardware design and verification \cite{intel_bug}. In a typical manual debug workflow, engineers first craft testbenches to expose bugs, simulate the design to reproduce them \cite{verilator}, analyze waveforms or assertions to pinpoint the root cause \cite{Zhang_Asgar_Horowitz_2022,vasudevan2021learning,5457129}, and finally patch the code. For complex systems, this trial-and-error cycle may require many iterations of simulation and manual inspection, making RTL debugging both time‑consuming and expensive.

Automatic Program Repair (APR)\cite{Fan_Gao_Mirchev_Roychoudhury_Tan_2023,Xia_Wei_Zhang_2023} has been proven effective in streamlining debugging for both software and hardware. There are two major ways for RTL-APR. 
Traditional techniques\cite{cirfix,rtl-repair,strider} typically combine program synthesis with symbolic execution to search in a predefined solution space, like AST manipulations. 
Recently emerged LLM-based methods\cite{Ahmad_Thakur_Tan_Karri_Pearce_2024b,hdldebugger,fixrag,rtlfixer,uvllm,location,Elnaggar,Fan_Gao_Mirchev_Roychoudhury_Tan_2023,Xia_Wei_Zhang_2023,origen} rely on learned statistics of RTL designs and natural language to directly generate patches in token space.

As shown in \autoref{fig:category}, we observed a spectrum of program repair operations that aligns with different classes of repair techniques, and a fix may be composed of multiple operations, namely \textbf{multi-step heterogeneity}. From top to bottom is the level of abstractions, descending from high-level design intent to low-level implementation detail, while the horizontal axis indicates the increasing difficulty of identifying a correct repair operation. The natural design flow proceeds from top to bottom: programmers first capture design intent in RTL, which is subsequently synthesized into netlists and simulated to produce waveforms. High-level operations rooted in design intent or RTL source code are inherently more amenable to LLM-based reasoning, whereas low-level operations involving netlists and dense waveforms are better suited to symbolic methods. Certain defects reside at the extreme high end of the spectrum, e.g., algorithmic errors, and fall beyond the scope of existing APR techniques. No fully automatic method can currently address them with reliable efficiency.

This observation raises a central challenge: \textbf{there is no silver bullet, no universal solution in RTL APR}. Specifications, interestingly, are only provided at the two extremes of the abstraction hierarchy, and any given APR technique works with one of them. A low-level specification precisely defines waveform segments via assertions or reference model comparison, while a high-level specification expresses design intent in natural language.
\textbf{Symbolic methods construct a search space from lower-level primitives}, e.g., fix a Boolean expression by constructing a sum-of-product expression encoding all possible combinations as an SMT formulation\cite{rtl-repair}. As the abstraction level rises, the search space expands exponentially and quickly becomes intractable, forcing these methods to degrade into less formal search engines guided by hand-crafted heuristics\cite{cirfix}.
On the other hand, \textbf{LLM-based methods draw on higher-level information}, such as the natural language specification and RTL source code. Their understanding of design intention and experience learned from prior designs enables them to rapidly resolve issues like coding style violations (e.g., mixing blocking and non-blocking assignments) or syntax errors. Yet low-level information, particularly dense waveform traces,  quickly saturates their limited context windows, inducing distraction and promoting hallucination. Most critically, \textbf{a practical fix often demands a coordinated sequence of diversified repair operations}, compounding the overall complexity.

Recent developments in LLM agents offer a unique opportunity to orchestrate LLM-based and symbolic APR approaches within a unified framework, capitalizing on the complementary strengths of both. The LLM agent paradigm enables an LLM to interact with an environment and iteratively refine its output. \textbf{Harness engineering}\cite{harness} has emerged as a systematic approach to constructing scaffold systems that equip agents for domain-specific tasks, integrating context engineering\cite{context-engineering}, guiding prompts, domain-specific skills, and a curated toolbox. In the context of RTL APR, harness engineering addresses how an LLM can effectively ingest complex RTL code and waveform data, invoke intricate linters and simulators, and distill overwhelming tool output without corrupting the context.  Moreover, a harness mitigates a critical concern in hardware verification: the inherent stochasticity of LLM outputs. The statistical nature and sampling-driven generation of LLMs yield unpredictable results, a challenge that can be tempered by validating outputs within the harness or by employing test-time scaling strategies\cite{test-time-scaling, self-consist,s*, tree-of-thoughts, CMCTS}.

In this paper, we introduce \textbf{Clover}, a neural-symbolic agentic harness for verified RTL repair. The framework orchestrates RTL APR as a structured search over code manipulations to explore a validated solution for the bug, performed by the main agent. Recognizing that distinct repair operations demand distinct techniques, the main agent dynamically delegates specialized subtasks to dedicated sub-agents or to symbolic solvers. We integrate and extend the SMT-based symbolic repair techniques from \cite{rtl-repair}, augmenting them with additional templates. The symbolic methods can precisely target their corresponding repair operations, thereby shielding the agent from low-level detail. As a component of the harness, we also introduce an RTL-specific toolbox that streamlines LLM–environment interaction. Furthermore, we propose a \textbf{Stochastic Tree-of-Thoughts} test-time-scaling strategy to govern the main agent. This mechanism records and samples dialogue states, allowing the agent to backtrack and explore alternative solution paths. Departing from the original Tree-of-Thoughts framework\cite{tree-of-thoughts}, we incorporate a heuristic distribution for expansion selection to balance exploration and exploitation, achieving faster and more reliable repair.

Our major contributions are:

\begin{itemize}
    \item \textbf{Clover harness}: We propose an agentic harness to automate RTL program repair with LLM-agents and symbolic approaches. A toolbox is designed to adapt the agent to RTL debugging.
    \item \textbf{Stochastic Tree of Thoughts test-time-scaling:} We propose a searching agentic workflow with a stochastic tree-of-thoughts sampling method, turning the LLM’s randomness into reliable search and consistently producing valid fixes.
    \item \textbf{Neural-symbolic Integration:} We orchestrate multiple task-specific LLM agent workflows together with SMT-based symbolic repairing to deal with diversified program repair operations. 
   
\end{itemize}

Evaluated on the RTL-Repair benchmarks \cite{rtl-repair}, experiments show that our approach can fix 96.8\% of bugs in the RTL-repair dataset within a given time limit, which covers 94\% and 63\% more bugs than traditional methods and other LLM-based approaches, while achieving an 87.5\% pass@1 rate on average, showing a high reliability.

\section{Background and Motivation}

\begin{table}
\caption{Comparison between different APR methods}
\label{tab:motivation}
\centering
\resizebox{\columnwidth}{!}{
\begin{tblr}{
  colspec={Q[m]Q[m]Q[m]Q[m]},
  cells = {c},
  vline{2} = {-}{},
  hline{2} = {-}{},
  hline{6} = {-}{},
  hline{9} = {-}{},
}
{Name}                        & {Fault\\Localization}     & {Patch\\Generation}                  & {Fix\\Capability} \\
CirFix\cite{cirfix}             & no                                & synthesis                                       & {limited}                                  \\
RTL-Repair\cite{rtl-repair}     & no                                & solving                                      & {limited}    \\
Strider\cite{strider}           & ~tracing                            & synthesis                                      & {limited}   \\
{\cite{Wu_Zhang_Yang_Meng_He_Mao_Lei_2022}} & {statistics}                    & {no}                                       & {no}                &                              \\
{Location-is-Key\\\cite{location}}  & experience                        & {experience\\translation}                   & {weak}                  \\
{RTLFixer\cite{rtlfixer}}         & {analysis\\experience}                       & {experience\\translation\\trial-and-error}                   & {weak}                        \\
{UVLLM\cite{uvllm}}               & {analysis \\tracing \\experience} & {experience\\translation}                   & {medium}                        \\
{Clover\\(this work)}              & {analysis \\experience}          & {experience\\trial-and-error\\solving}               & {strong}     
\end{tblr}
}
\end{table}
\begin{figure*}[t]
    \centering
    \includegraphics[width=0.8\textwidth]{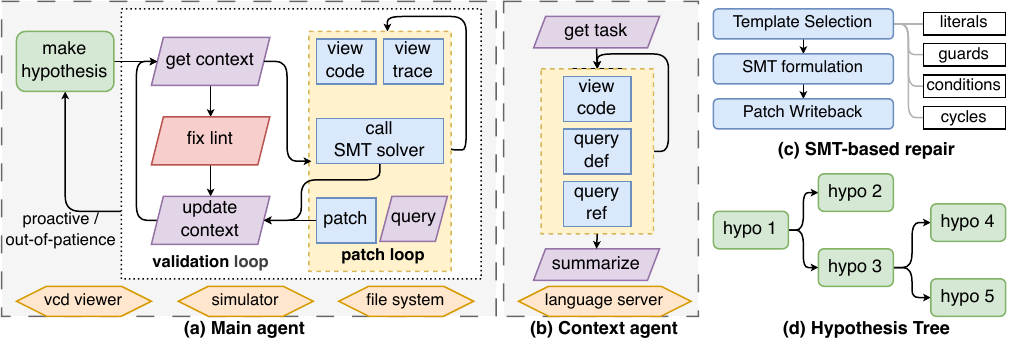}
    \caption{Overview of the \rdba~Framework}
    \label{fig:overview}
\end{figure*}

\subsection{RTL Debugging}

RTL debugging aims to restore correct functionality with minimal code modifications and typically proceeds in three stages: bug discovery, fault localization, and patch generation.  
Bug discovery identifies a test case that exposes the fault, either through simulation\cite{verilator} or verification methods\cite{vasudevan2021learning,5457129}. Fault localization then narrows the suspect code region\cite{Wu_Zhang_Yang_Meng_He_Mao_Lei_2022,location, strider}, and patch generation resolves the fault while minimizing edits\cite{cirfix,rtl-repair,strider, uvllm}. 



This paper targets APR, which assumes bugs are known and focuses on the latter two stages. 
We define correctness by comparing waveforms during simulation, a method that is both comprehensive for LLMs and standard in RTL debugging. A setting with minimal human guidance is assumed, as in traditional methods: \textbf{no design specification in natural language and the LLM observes only the source code and the I/O waveforms of the top module}.

\subsection{Program Repair}


Program repair aims to modify a program within a bounded editorial distance such that it passes a given test. 
\autoref{fig:category} summarizes a representative spectrum of common repair operations. 
\textit{Per spec} refers to aligning code behavior with a natural language specification. 
\textit{Coding style} restructures RTL code to conform to standard conventions (e.g., restricting sensitivity lists to \texttt{posedge clock}). 
\textit{Add signal dependency} introduces a new dependency to an existing signal. 
\textit{Adjust submodule} instantiates or removes submodules. 
\textit{Algorithm} addresses algorithmic flaws that may require substantial changes to the logic. 
\textit{Syntax} corrects RTL syntax errors. 
\textit{Handshake} resolves issues on handshaking interfaces. 
\textit{Control flow} fixes control constructs such as \texttt{if-else} or \texttt{generate for}. 
\textit{Cycle shift} adjusts signal timing across cycles by inserting or removing pipeline registers. 
\textit{Word-expr} and \textit{bool-expr} reshape word-level or bit-level expressions without adding new dependencies. 
\textit{Literals} supplies a new constant value for a static signal.

\subsection{SMT-based Symbolic Repair}

RTL-repair \cite{rtl-repair} introduces a symbolic repair approach that formulates a subset of repair operations as an SMT problem. The hardware design is synthesized into a transition model under bounded model checking, with waveform specifications expressed as assertions on inputs and outputs; repair operations are then encoded as free variables. Due to the inherent complexity, the method relies on predefined templates that target specific repair types. Three templates are proposed: \textit{replace literal}, \textit{add guard}, and \textit{conditional overwrite}, which correspond respectively to the \textit{literals}, \textit{bool-expr}, and (partially) \textit{word-expr} repair categories shown in \autoref{fig:category}.

\subsection{Motivation}

We summarize existing approaches for fault localization and patch generation, along with their fix capability, in \autoref{tab:motivation}. 
Regarding fault localization, \cite{cirfix} and \cite{rtl-repair} bypass this step entirely. 
\cite{strider} and \cite{uvllm} trace mismatched signals, a technique that, while powerful, is often impractical due to its reliance on internal signal waveforms. 
\cite{Wu_Zhang_Yang_Meng_He_Mao_Lei_2022} adopts spectrum-based fault localization borrowed from software debugging. 
LLM-based methods, in contrast, leverage learned patterns to infer fault locations, and some \cite{rtlfixer, uvllm} incorporate static checks from linters \cite{verilator, iverilog} for additional guidance.

Fix capability is intrinsically tied to the underlying patch generation strategy. 
We define capability as the probability of producing a valid repair within prescribed time and resource constraints. 
\cite{cirfix} and \cite{strider} employ program synthesis, while \cite{rtl-repair} reduces the task to SMT solving. 
In both cases, the fix capability is confined to a set of predefined templates. 
LLM-based methods eliminate such template restrictions, yet they struggle with low-level hardware details due to limited context windows and high computational cost under frequent invocation. 
Many of these approaches translate specifications to code directly, a paradigm ill-suited for large designs where sufficiently detailed specifications are rarely available. 
Some efforts \cite{location} explore fine-tuning to improve outcomes.
Notably, \cite{uvllm} strengthens its performance by incorporating the signal-tracing strategy of \cite{strider}.


Addressing the challenge of \textbf{multi-step heterogeneity}, neither purely symbolic nor purely LLM-based approaches can handle the full spectrum of APR operations alone. 
A natural resolution is to combine them into a neural-symbolic framework. 
Given the rigidity of symbolic methods, which either enforce exact compliance with the specification or fail completely, LLMs are best positioned as high-level orchestrators. 
They can first resolve problems at a more abstract granularity and then delegate the final, low-level adjustments to symbolic techniques. 
Such a system can be constructed as a harness for the RTL APR agent, exposing symbolic tools and RTL-specific utilities as an environment. 
By applying test-time scaling to mitigate LLM stochasticity, this design enables reliable, verifiable repair.

\section{Overview}

\autoref{fig:overview} illustrates the core components of the \rdba~framework. 
\autoref{fig:overview}(a) depicts the workflow of the main agent, which orchestrates all sub-agents and manages symbolic repair. 
Two sub-agents operate under the main agent's control: a context agent and a lint-fix agent. 
These sub-agents are designed with narrow, task-specific responsibilities to reduce the cognitive load on the main agent and maintain a clean LLM context. 
The context agent, shown in \autoref{fig:overview}(b), extracts critical code fragments from the extensive RTL codebase. 
The lint-fix agent follows a similar workflow but targets errors and warnings reported by linters. 
A detailed discussion of these agents as a harness appears in \autoref{sect:harness}.

The SMT-based symbolic repair technique functions as a tool invoked by the main agent when a predefined repair template is deemed applicable. 
This approach is inspired by RTL-repair \cite{rtl-repair} and extended with an additional template, adapted for integration with the LLM agent, as outlined in \autoref{fig:overview}(c) and elaborated in \autoref{sect:symbolic}. 
To further enhance reliability, we introduce the Stochastic Tree-of-Thoughts test-time scaling method, which enables the main agent to search over a tree of hypotheses, as shown in \autoref{fig:overview}(d) and detailed in \autoref{sect:tot}.

\section{Multi-Agent Harness}
\label{sect:harness}

\subsection{The main agent}

As illustrated in \autoref{fig:overview}(a), the main agent orchestrates the overall APR process. 
Its workflow is structured as a three-level nested loop: an outer loop for hypothesis generation, a middle validation loop for hypothesis verification, and an innermost patch loop for concrete patch formation.

The hypothesis-making loop emulates the trial-and-error reasoning of a human programmer. 
When confronted with a bug, a programmer typically posits a hypothesis regarding the root cause based on observable symptoms, then validates it by applying and testing a patch. 
Through this iterative cycle, the programmer refines their understanding and may formulate a new hypothesis if the current one proves incorrect. 
In our workflow, the agent either proactively proposes a new hypothesis or triggers an out-of-patience mechanism after exceeding a predefined operation budget. 
This simple mechanism curtails resource expenditure on any single hypothesis, encouraging broader exploration of the solution space.

The validation loop begins by retrieving relevant code context from the context agent. 
Depending on whether linter errors or warnings are present, the main agent either invokes the lint-fix agent or proceeds directly to the patch loop to synthesize a patch. 
Once a patch is generated, it is registered with the context agent and submitted to the simulator for verification. 
In summary, this loop primarily delegates subtasks and applies exactly one patch per iteration. 
Patches are applied cumulatively to enable multi-step repair, such as using LLMs to introduce new signal dependencies before refining a Boolean expression via the SMT solver.

The patch loop gathers the necessary information to construct a concrete patch. 
In each iteration, the LLM may perform one action: it can inspect code snippets or trace files, or query the context agent for additional details. 
This process continues until the LLM either determines that a direct patch can be emitted or concludes that invoking the SMT solver with a specific repair template is appropriate.

\subsection{Multi-agent Interaction}

The main agent delegates specific subtasks to subordinate agents, thereby maintaining a focused context, accessing specialized tool sets, and executing task-specific workflows. 
The orchestration involves two sub-agents: a context agent and a lint-fix agent.

The context agent is responsible for gathering critical code fragments needed to validate a hypothesis. 
For each hypothesis proposed by the main agent, a context agent instance is created, maintaining a continuous LLM dialogue across multiple patches in the validation loop. 
As shown in \autoref{fig:overview}(b), each invocation of the context agent begins with a task from the main agent, performs a series of navigation actions within the RTL codebase, and returns a summarized response. 
A task may consist of receiving a newly applied patch or answering an explicit query from the main agent. 
Equipped with a language server, the context agent traces signal dependencies and module hierarchies using \textit{query def} to locate symbol definitions and \textit{query ref} to identify all reference sites.

The lint-fix agent addresses errors and warnings from the linter, or determines that a given warning should be suppressed. 
Its workflow is straightforward: gather relevant context and then decide on an appropriate fix or ignore action. 
We separate this agent from the main agent because lint resolution, while trivial for LLMs, is ubiquitous during RTL APR. 
Upon receiving a lint message, the main agent spawns a new lint-fix agent to obtain a patch, thereby insulating the main dialogue from low-level linting details and preserving a clean context.

\subsection{RTL-specific Tool Usage}

As depicted in \autoref{fig:overview}(a) and (b), the agents are provisioned with an RTL-specific toolbox. 
Basic file reading and editing are performed via the file system. 
For signal traces (e.g., VCD files), we provide a VCD viewer that returns textual signal traces over a specified time window, with deviations from the golden reference explicitly highlighted and optimizations such as suppressing wide-bit signals. 
We further simplify environment interaction by allowing the agent to invoke the RTL simulator and retrieve feedback with minimal friction. 
Additionally, we integrate the slang-server \cite{slang-server} as an RTL language server. 
Language servers serve as the backend for IDEs such as VS Code, enabling code navigation and linting through parsed program representations. 
For linting, we primarily rely on Verilator \cite{verilator}, supplemented by a custom linter that flags constructs that are legal under the Verilog specification yet error-prone in practice (e.g., multiply driven signals or partially driven wires).

\section{SMT-based Symbolic Repair}
\label{sect:symbolic}


Our repair module extends RTL-Repair \cite{rtl-repair} with additional repair templates and integrates them within the LLM agent framework. 
The process is illustrated in \autoref{fig:overview}(c). 
Rather than applying all templates in a fixed sequence, the main agent selects a single repair template that best aligns with the inferred bug mechanism. 
The chosen template is then formulated as an SMT problem: the RTL code is translated into a bounded model checking instance, and repair options are encoded as free variables. 
After the SMT solver produces a solution, we convert the symbolic outcome into structured repair actions instead of directly modifying the source code. 
The agent then synthesizes the final source-level patch, ensuring minimal disruption and respect for the original coding style.


\subsection{Cycle Shift Templates}
\label{sec:temporal_templates}

\begin{figure}
    \centering
    \includegraphics{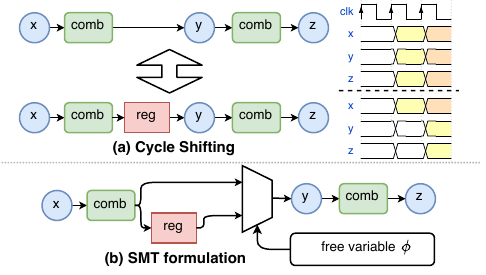}
    \caption{Cycle Shifting and Its SMT formulation}
    \label{fig:cycleshift}
\end{figure}

To address temporal RTL bugs, we introduce an additional template that explicitly models cross-cycle behavior. 
As shown in \autoref{fig:cycleshift}(a), a signal \textit{y} can be either a wire or the output of a register, determining whether \textit{y} changes concurrently with \textit{x} or is delayed by one cycle. 
To capture this cycle-shifting behavior in SMT, we introduce a free variable $\phi$ that serves as the selection signal for a multiplexer, as depicted in \autoref{fig:cycleshift}(b). 
The SMT solver then assigns a constant value to $\phi$ such that the output satisfies the bounded model checking assertions.

Indiscriminately inserting such multiplexers on all signals is hazardous, as it can introduce combinational loops. 
In a purely symbolic approach, this template would be infeasible because the solver cannot determine the appropriate signals to modify. 
With the assistance of the main agent, however, the template can be selectively applied to a targeted set of signals identified by the agent.

\subsection{Agentic Template Selection and Repair}
\label{sec:agent_template_patch}

A key enhancement afforded by the agent framework is that template-based repair ceases to be an isolated backend pass and instead becomes a context-aware decision guided by the debugging process. 
Whereas the original RTL-Repair workflow attempts its three templates sequentially until one yields a satisfying repair, our approach delegates template selection to the main agent, providing it with detailed instructions regarding the effect and applicability of each template.

After SMT solving, Clover further diverges from prior flows in how the repair is materialized. 
The backend returns a structured repair result rather than directly emitting a patch. 
This result identifies the selected template and normalizes the synthesized repair into explicit source-level actions, e.g., expression rewrites, guarded insertions, or temporal-structure modifications. 
The patch-generation agent then interprets these actions under template-specific constraints and emits the final patch while preserving the original control-flow structure, statement ordering, and coding style. 
For instance, when rewriting the conditions of a signal, the agent may locate a suitable \texttt{if-else} branch to apply the modification, avoiding the introduction of an explicit multiplexer tree.

Furthermore, compared with the original RTL-Repair implementation, Clover extends this agent-mediated template repair flow to System Verilog designs, thereby broadening the applicability of the symbolic repair strategy to a wider range of modern RTL codebases.
\section{Stochastic Tree-of-Thoughts}
\label{sect:tot}


\subsection{Hypothesis Tree}

As shown in \autoref{fig:overview}(d), the hypotheses proposed by the main agent are organized into a tree structure, which substantially enhances the capacity to explore diverse solutions to a given bug. 
This mechanism is realized by applying our Stochastic Tree-of-Thoughts test-time scaling algorithm to the main agent. 
Test-time scaling \cite{test-time-scaling} refers to the allocation of additional computational resources during LLM inference to improve output quality. 
Tree-of-thoughts, a specific test-time scaling technique, structures LLM inferences as a search tree. 
Within this hypothesis tree, each node can branch into multiple child nodes, corresponding to new hypotheses formulated based on observations gathered while validating the parent hypothesis. 
The original tree-of-thoughts algorithm employs depth-first or breadth-first search combined with beam search, and relies on the LLM itself to evaluate each node, leading to blind exploration over a near-infinite space. 
In contrast, our approach introduces an auxiliary sampling procedure to dynamically balance the trade-off between exploration and exploitation within the tree, guided by a heuristic function that evaluates the promise of each node.

\subsection{Heuristic Function and Algorithm}\label{subsec:value_func}

Instead of asking LLM to evaluate a node, we provide a heuristic value function for stable outcomes.
However, it is hard to precisely determine how far a given node is from the goal, since the location of the target region is unknown. We define a heuristic function $f(c,h)$, where $c$ is the code status and $h$ is the LLM dialogue history. It estimates the likelihood that the search can succeed from a node $(c,h)$ based on prior knowledge, as shown by \autoref{equ:f}. The meaning of symbols is defined in \autoref{tab:symbol}. 
This function encourages passing more test benches $tb_p(c)$ and getting more information $N_Q(h)$ from the design, while avoiding compilation errors $N_{CE}(h)$, excessive patches $N_p(h)$, and token usage $N_{tok}(h)$. The base value $b$ adjusts the probability to balance between exploiting good nodes and exploring temporarily worse nodes. These features can be gathered during the validation loop in the main agent. The coefficients $\lambda$ balance the strength of each factor.
As a test-time scaling approach, the coefficients can be changed without the cost of retraining.

Furthermore, we choose the state to expand stochastically to balance exploration and exploitation.
In each search step, the probability of choosing a state $s=
(c,h)$ to expand is defined in \autoref{equ:softmax}. The algorithm is simple: in each step, a node is sampled from the distribution, then its code status and dialogue history are restored for the main agent. The main agent runs until it proposes a new hypothesis, which is added to the nodes for the next sampling step. 

\begin{align}
    f(c,h) &= \lambda_1\cdot \frac{tb_p(c)}{N_{tb}}+\lambda_2\cdot N_Q(h)-\lambda_3\cdot N_{CE}(h)\nonumber\\
    &-\lambda_4\cdot N_{tok}(h)-\lambda_5\cdot N_P(h)+b \label{equ:f}\\
    Pr[s&=(c,h)] = \frac{e^{f(c,h)}}{\sum_i e^{f(c_i, h_i)}}\label{equ:softmax}
\end{align}

\begin{table}[t]
\centering
\caption{Symbols in the heuristic functions}
\label{tab:symbol}
\resizebox{0.8\columnwidth}{!}{
\footnotesize
\begin{tblr}{
  row{1} = {c},
  vline{2} = {1}{},
  vline{2} = {2-14}{},
  hline{2} = {-}{},
}
Name                              & Meaning                        \\
$tb_p(c)$                & {Number of passed testbenches}      \\
$N_{tb}$                 & Number of testbenches                \\
$N_Q(h)$                   & Number of queries                    \\
$N_{CE}(h)$              & {Number of unsolved compile errors} \\
$N_{tok}(h)$             & Number of used tokens                \\
$N_P(h)$                   & Number of patches                    \\
$b$                         & Base value                           \\
$\lambda_i$  & Coefficients
\end{tblr}
}
\end{table}

\section{Experiments} \label{sec:experiment}

\begin{figure*}[t]
\centering
\includegraphics[width=\linewidth]{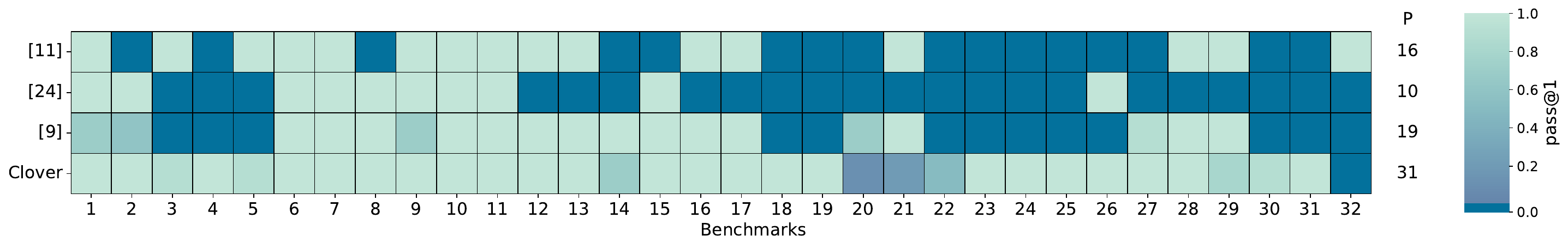}
\footnotesize
Benchmarks: 1. decoder\_w1 2. decoder\_w2 3. counter\_k1 4. counter\_w1 5.counter\_w2 6. flop\_w1 7. flop\_w2 8. fsm\_w1 9. fsm\_w2 10.fsm\_s1\\
11.fsm\_s2 12.shift\_w1 13.shift\_w2 14.shift\_k1 15.mux\_k1 16.mux\_w1 17.mux\_w2 18.sha3\_r1 19.sha3\_w1 20.sha3\_w2
21.sha3\_s1\\ 22.pairing\_w1 23.pairing\_w2 24.pairing\_k1 25.reed\_b1 26.reed\_o1 27.sdram\_w1 28.sdram\_w2 29.sdram\_k2 30.i2c\_w1 31.i2c\_w2 32.i2c\_k1
\caption{Evaluation on Bug Repairing Ability for RTL-repair\cite{rtl-repair}, MEIC\cite{MEIC}, UVLLM\cite{uvllm} and Clover. \\The first 17 benchmarks are easy
cases with only one module and one file, and the rest are much more complex. The color implies the pass@1 rate of each case. P: total passed tests. 
}
\label{fig:fix}
\end{figure*}

\subsection{Experiment Setup}



We evaluate using the RTL-repair\cite{rtl-repair} dataset, which contains a variety of buggy RTL designs, widely used in prior works\cite{rtl-repair,cirfix}. The roots of bugs varied from structural to algorithmic. Our results are compared against three baselines. For traditional APR, \textit{RTL-repair}\cite{rtl-repair} is the state-of-the-art, which transforms program repair into SMT problems. For LLM-based APR designed for RTL, MEIC\cite{MEIC} and UVLLM\cite{uvllm} are recent works that are open-sourced. In our evaluation, a case is passed if the waveform matches the golden reference, the same as prior works. Bug discovery or testbench generation is a different research domain.




We evaluate our approach with API calls from the Seed2-code model. Only the source code is modifiable to avoid the agent removing the tests. We use Verilator\cite{verilator} to lint and simulate the designs, and build additional linting on top of Yosys \cite{yosys} processed net list. The patched code is verified by simulation to ensure correctness on the given testbenches.
To estimate the reliability, we use the pass@k metric, which means the probability that the bug is fixed within k independent retries within the given time and resource limit. 
We sample pass@k with 10 retries, each with a 30-minute timeout and 2M total token limit. Because token usage can only be checked after the LLM finishes inference, the actual time and token usage may be slightly more than the limit. The baselines do not support such a limit, so we adjust the number of rounds and retries for a similar time usage. The coefficients in \autoref{equ:f} are $\lambda_1=50, \lambda_2=1, \lambda_3=5, \lambda_4=0.0005, \lambda_5=3, b=10$.


\subsection{Bug Repair Results}

We first evaluate the ability to reliably fix a bug within the retry, time, and resource limit. As shown in \autoref{fig:fix}, our approach can fix 31 of the bugs, while \textit{RTL-repair}, \textit{MEIC}, and \textit{UVLLM} can only fix 16, 12, and 20 of them, respectively. 

The first 17 benchmarks are easy cases with only one module and one file, and the LLM can fix them easily. However, some of the designs have some differences from commonly used textbook designs, e.g., the 3-to-8 decoder has an enable signal that is not useful. This causes the agent to be confused at first, but it can be recovered after validating the agent's understanding by trial-and-error. 

The latter 15 benchmarks are much more complex. Their root cause is hidden deep inside the module tree and file structure. 
Some faults can be found by linters. \textbf{sha3\_r1}, \textbf{pairing\_k1}, and \textbf{reed\_b1} have incorrect width on signals or expressions, which can trigger warnings in Verilator. \textbf{sha3\_w1} has a wire partially undriven, and \textbf{pairing\_w2} mistakenly exchanges the input and output of a module. Though legal in Verilog, these error-prone cases can be detected by our additional linter. 
Others can benefit from the information collected by interaction: 
The dead loop in \textbf{pairing\_w1} is an algorithmic error that changes the direction of a loop variable, which is suspicious and can be identified. 
\textbf{sha3\_w2} miswrites a condition into 0 and changes a wire into a register.
\textbf{sha3\_s1} drops an update condition when the buffer is full. These two cases are rather difficult as they combine multiple repair operations. Our approach still managed to search for a solution. We fail to fix the \textbf{i2c\_k1} case. \textbf{i2c} benchmarks require simulating in-cycle latency and z-value, which is a different category of RTL design, and our tools do not support it well.

For \textit{RTL-repair}, many bugs fall out of its predefined templates. For example, its fix range cannot cover the bug of \textbf{decoder\_w2}, \textbf{fsm\_w1}, \textbf{counter\_w1}, and \textbf{shift\_k1} benchmarks has some wrong "if-else" or "case" structures or wrong sensitivity lists in an "always" block. Though \textit{RTL-repair} passes the \textbf{i2c\_k1} case, we found that it manually instructed the repairer to fix the one file where the bug is located to reduce the scale of the SMT problem, which is forbidden in our minimal human guidance setting.  
\textit{MEIC} ranks RTL codes with a scorer agent, which further increases the instability.  \textit{UVLLM} localizes fault by Verilator linting and Strider signal tracing, and samples multiple LLM outputs from the same input. This strategy works well in smaller cases, but only linting works in more complex multi-file cases. Besides, \textit{UVLLM} requires a problem description text as input, but our setting assumes no specification. If a specification is provided, then it can pass the simple cases of \textbf{counter}, but has no improvement in other cases. For the \textbf{counter} cases, signal tracing fails to trace through combinational logics, and patch by experience also fails because the design is uncommon: the "overflow" signal is not related to the "enable" signal. Without specification, the LLM is often hallucinated to overwrite the design. Our approach prevents such a hazard by verifying the hypotheses and reviewing the feedback.

We found that the original testbench in RTL-repair for the "bug" of \textbf{reed\_o1} has no effect. We keep it unchanged for a fair comparison with RTL-repair.

In conclusion, our approach can reliably fix the bugs within the time limit, passing 63\% and 94\% more bugs than LLM-based\cite{MEIC, uvllm} and traditional methods\cite{rtl-repair}, respectively. \textit{RTL-repair} has limited fix capability, while \textit{MEIC} and \textit{UVLLM} do not address the characteristics of LLM for a reliable outcome. We achieved 87.5\% pass@1 rate, showing high reliability. Our average time usage is 413.8 seconds (241.8 if failed cases are excluded), and our average token usage is 294.8k.


\subsection{Ablation Study}\label{subsec:reliable}

\begin{figure}[t]
    \centering
    \includegraphics[width=0.9\columnwidth]{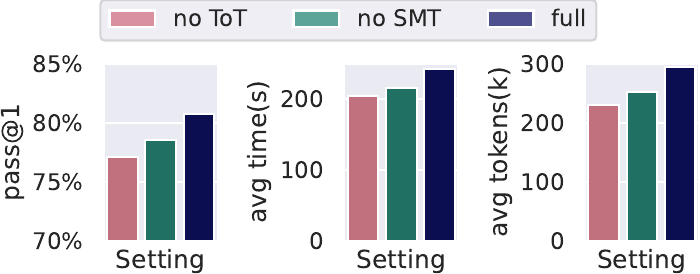}
    \caption{Ablation study on different settings on the complex benchmarks (\#18-\#32 in \autoref{fig:fix}). The average time and tokens exclude failed trials. }
    \label{fig:ablation}
\end{figure}

We evaluate the effectiveness of our SMT-based repair module and the Stochastic Tree-of-Thoughts method. 
As shown in \autoref{fig:ablation}, we compare three configurations: \textit{no ToT} disables Stochastic Tree-of-Thoughts on the main agent, which then iteratively proposes new hypotheses based solely on the immediately preceding one; \textit{no SMT} prevents the main agent from invoking SMT-based repair, reducing it to a pure LLM agent; and \textit{full} enables all proposed techniques. 
Evaluation is conducted on the more challenging benchmarks (\#18-\#32 in \autoref{fig:fix}), as the baseline capabilities of LLMs readily solve the simpler cases, rendering comparisons uninformative. 
We report pass@1 rate, average execution time, and average token usage. 
Failed trials are excluded from the time and token averages because successful runs consume substantially fewer resources than the imposed limits (successful trials average 241 seconds, compared to a 1800-second timeout).

As shown in \autoref{fig:ablation}, the \textit{no ToT}, \textit{no SMT}, and \textit{full} configurations exhibit a consistent increasing trend across all three metrics. 
\textit{no ToT} yields the lowest pass@1 rate, underscoring the importance of exploring diverse root-cause hypotheses for complex bugs, as a strictly linear hypothesis chain frequently terminates in dead ends. 
Although expanding the hypothesis tree incurs additional time and token costs, the overhead remains within acceptable bounds. 
The \textit{no SMT} configuration also trails \textit{full} in pass@1 rate, indicating that SMT-based repair contributes to stabilization in specific cases. 
\textit{full} consumes more time and tokens due to the overhead of template selection and post-solving patch synthesis by the main agent. 
The improvement in pass@1 is modest, primarily because the selected benchmarks are not sufficiently difficult and the baseline pass rate already exceeds 75\%.

In summary, the integration of neural-symbolic techniques with Stochastic Tree-of-Thoughts enhances bug-fixing capability at a modest increase in time and token consumption.

\subsection{Synthetic Benchmarks Results}\label{subsec:generalize}

\begin{table} 
\centering
\caption{Evaluation on the Synthetic Benchmarks.}
\label{tab:synth}
\resizebox{\columnwidth}{!}{
\begin{tblr}{
  vline{2} = {2-9}{},
  hline{2} = {-}{},
}
Benchmark & Error                         & Result    & Pass@1 \\
synth\_s1 & Incorrect binary operator     & \ding{51} & 80\% \\
synth\_s2 & Duplicated item in expression & \ding{51} & 100\% \\
synth\_s3 & Negate if condition           & \ding{51} & 100\% \\
synth\_s4 & Additional minus one          & \ding{51} & 100\% \\
synth\_s5 & Missing item in expression    & \ding{51} & 100\% \\
synth\_s6 & Incorrect reduce operator     & \ding{51} & 80\% \\
synth\_s7 & Delayed by one cycle          & \ding{51} & 100\%\\
synth\_s8 & Advanced by one cycle         & \ding{51} & 100\%
\end{tblr}
}
\end{table}

The RTL-repair benchmarks consist of common hardware modules, which may have open-source implementations and thus could appear in an LLM’s training data. To test generalization, we design a synthetic benchmark with purely random logic that carries no semantic meaning or no practical usage. The synthetic module contains randomly wired signals and operators, with simple internal states and logic. We insert a variety of bugs into this synthetic design. The errors and results are shown in \autoref{tab:synth}. This ensures these benchmarks did not appear in training data. SMT-based repair is disabled for this experiment. The results show that, even without meaningful semantics, the LLM-based approach can still fix the RTL bugs by analyzing error symptoms, guessing likely problematic locations, and deducing the correct code from waveform comparison, demonstrating that our approach can fix bugs in completely new designs. 


\textbf{synth\_s1} and \textbf{synth\_s6} require the agent to guess the correct operator with very little context—essentially a program synthesis task—resulting in low success rates. \textbf{synth\_s2}, \textbf{synth\_s3}, \textbf{synth\_s4}, and \textbf{synth\_s5} contain combinational logic errors that could be inferred from the waveform trace, showing the agent’s capability for simple logical reasoning. \textbf{synth\_s5} is harder as it requires introducing new dependencies and needs more guesses. \textbf{synth\_s7} and \textbf{synth\_s8} need to remove or add a register instance. Thanks to the ADI waveform dumping functionality, the agent could observe in the trace when a signal is delayed or advanced by one cycle, enabling it to repair these two types of bugs. 


\section{Conclusion}

RTL APR faces the multi-step heterogeneity challenge, and neither symbolic nor LLM-based APR methods alone can cover all repair operations. We propose \rdba, a neural-symbolic agentic harness with stochastic Tree-of-thoughts. By exploring a hypothesis tree while applying both SMT- and LLM-based repairing, \rdba~reliably fixes 96.8\% of bugs and achieves an 87.5\% pass@1 rate on average in the RTL-repair benchmarks.

\newpage

\bibliographystyle{ACM-Reference-Format}
\bibliography{bib.bib}

\end{document}